# Crystal Symmetry Breaking in Few-Quintuple $Bi_2Te_3$ Films: Applications in Nanometrology of Topological Insulators

K. M. F. Shahil, M. Z. Hossain, D. Teweldebrhan and A. A. Balandin*

Nano-Device Laboratory, Department of Electrical Engineering and Materials Science and Engineering Program, Bourns College of Engineering, University of California – Riverside, Riverside, California 92521 USA

(Submitted to Applied Physics Letters, February 2010)

**Abstract**

The authors report results of micro-Raman spectroscopy investigation of mechanically exfoliated single-crystal bismuth telluride films with thickness ranging from a few-nm-range to bulk limit. It is found that the optical phonon mode $A_{1u}$, which is not-Raman active in bulk $Bi_2Te_3$ crystals, appears in the atomically-thin films due to crystal-symmetry breaking. The intensity ratios of the out-of-plane $A_{1u}$ and $A_{1g}$ modes to the in-plane $E_g$ mode grow with decreasing film thickness. The evolution of Raman signatures with the film thickness can be used for identification of $Bi_2Te_3$ crystals with the thickness of few-*quintuple* layers important for topological insulator and thermoelectric applications.

* The author to whom correspondence should be addressed. Electronic address: balandin@ee.ucr.edu ; group web-site: http://www.ndl.ee.ucr.edu

Topological insulators (TI) are materials with an insulating gap exhibiting quantum-Hall-like behavior in the absence of a magnetic field [1-6]. It was suggested that TIs can be used for realization of quantum computing because they contain surface states that are topologically protected against scattering by time-reversal symmetry. TI films were also proposed for applications in the magnetic memory where write and read operations are achieved by purely electric means. The bismuth telluride ($Bi_2Te_3$) family of materials was demonstrated to be TI with a large energy gap and surface states consisting of a single Dirac cone [5]. The single-crystal $Bi_2Te_3$ films with the thicknesses of just a few nanometers are particularly promising as TI materials both for investigating their unique physical properties and potential practical applications. The crystal structure of $Bi_2Te_3$ is rhombohedral with five atoms in one unit cell [7-10]. The lattice parameters of the hexagonal cells of $Bi_2Te_3$ are $a_H$ = 0.4384 nm and $c_H$ = 3.045 nm. Its atomic arrangement can be visualized in terms of the layered structure with each layer, referred to as a *quintuple*, consisting of five mono-atomic planes of -Te-Bi-Te-Bi-Te. The quintuple layers are weakly bound to each other by the van der Waals forces. While the band gap in a single quintuple is larger than in a few-quintuple layer (FQL) the latter has less coupling between the surface states of the top and bottom interfaces.

We have recently demonstrated "graphene-like" mechanical exfoliation of the atomically-thin single-crystal films [11] and ribbons [12] of $Bi_2Te_3$. Using a combination of the optical microscopy, atomic force microscopy (AFM) and scanning electron microscopy (SEM) we have shown that a bulk bismuth telluride crystal can be cleaved into films with the thickness down to ~1 nm, which corresponds to a single quintuple [11]. The exfoliation process was analogous to the one initially developed for graphene [13]. Taking the graphene analogy even further one should consider using micro-Raman spectroscopy as a nanometrology tool for identification of FQL films of $Bi_2Te_3$ and assessing their quality. Indeed, Raman spectroscopy has proven to be the most reliable tool for counting the number of atomic planes in graphene flakes via deconvolution of its 2D (G') band [14]. It has been used at different temperatures and on various substrates [15-16].

In this letter, we report the results of our micro-Raman investigation of the mechanically exfoliated single-crystal $Bi_2Te_3$ films with the thickness ranging from FQL to the bulk limit. We found that several interesting features in Raman spectra of few-nm-thick $Bi_2Te_3$ films create a possibility for using micro-Raman spectroscopy as a nanometrology tool for TIs. We focused on FQL because due to low thermal conductivity of $Bi_2Te_3$ a systematic Raman study of individual quintuples is complicated by strong local heating and melting even at low excitation power. Moreover, FQL are more practical for TI and thermoelectric applications.

The FQL samples for this study were prepared and identified using the same technique described by us elsewhere [11-12]. An example of an exfoliated FQL with the lateral dimension of ~ 9 μm × 3 μm and uniform thickness is shown in Fig. 1. A detail AFM inspection confirms a uniformity of this film with the thickness of ~15 quintuples. Some of the obtained FQLs were transferred to $Si/SO_2$ wafers with prefabricated trenches for better visualization and Raman characterization (see inset in Fig. 1). The thickness of different regions of the flakes was determined through AFM scans. In Fig. 2 we present the AFM height profile, which starts at zero height (substrate) goes to the thick (~40 nm) region and then falls back to the thin (~4 nm) region, which corresponds to a four-quintuple layer. The inset shows the actual scan area.

We used an upgraded Renishaw inVia microscope for this study. All spectra were excited at room temperature with laser light (λ = 488 nm) and recorded in backscattering configuration through a 50× objective. An 1800-lines/mm grating provided a spectral resolution of ~1 $cm^{-1}$, which was software-enhanced to ~ 0.5 $cm^{-1}$. Since $Bi_2Te_3$ has very low thermal conductivity of 1.5 W/mK (0.6 W/mK) along (perpendicular) the cleavage plane [17] and a low melting point, a special care was taken to avoid local heating and melting during the measurements. The optimum excitation power in our setup was determined to be ~0.25 mW on the surface. It provided good signal-to-noise (S/N) ratio without introducing damage to the samples. To improve S/N we accumulated spectra from several spots within the same thickness region and then averaged them.

Bulk $Bi_2Te_3$ is a semiconductor with a five-atom primitive cell, which belongs to the space group $R\overline{3}m(D_{3d}^5)$ [7]. Consequently, bulk crystals reveal 15 lattice vibration modes (phonon polarization branches). Three of these branches are acoustic and 12 are optical phonons. According to the group theory classification, 12 optical branches have $2A_{1g}$, $2E_g$, $2A_{1u}$, and $2E_u$ symmetry [8]. Due to the inversion symmetry of the crystal these phonon modes are exclusively either Raman or infrared (IR) active [8]. Fig. 3 (a) presents Raman spectra of bulk $Bi_2Te_3$ crystal and three FQLs as their thickness $H$ decreases from ~82 nm to ~4 nm. The frequencies of the observed peaks and their assignment are listed in Table I together with previously reported data for bulk $Bi_2Te_3$. One can see that our results for four Brillouin zone (BZ) center Raman active modes, $E_g^1$ (TO), $A^1_{1g}$ (LO), $E_g^2$ (TO), and $A^2_{1g}$ (LO), are consistent with literature [7-10]. In this nomenclature, TO and LO are transverse and longitudinal optical phonons, respectively. Both $E_g$ and $A_{1g}$ modes are two-fold degenerate: in $E_g$, the atoms vibrate in the basal plane, while in $A_{1g}$, the atoms vibrate along $c_{H.}$ [8]. The $A^1_{1g}$ and $Eg^1$ vibrations occur at lower frequencies than $A^2_{1g}$ and $E_g^2$. The latter modes, where the outer Bi and $Te^{(1)}$ atoms move in opposite phase, are mainly affected by the forces between Bi and $Te^{(1)}$ atoms (see Fig. 3b). In $E_g^1$ and $A^1_{1g}$ modes the outer Bi-Te $^{(1)}$ pairs move in phase. Thus the Bi-Te $^{(2)}$ bonding forces are primarily involved in these vibrations. In crystals with inversion symmetry, the IR-active modes ($A_{1u}$) must be odd parity while the Raman-active modes ($E_g$, $A_{1g}$) must be even parity under inversion [18]. The odd-parity phonons (IR active) do not show up in Raman spectra for bulk samples as long as crystal retains its symmetry.

It is interesting to notice in Fig. 3(a) that an additional peak appears at ~117 cm$^{-1}$ in FQLs. Its intensity, normalized to the intensity of $E_g^2$ (the most pronounced feature in the spectrum), grows with decreasing FQL thickness (see Table I). This peak was identified as $A_{1u}$ mode composed of TO phonons at the BZ boundary (Z point). The $A_{1u}$ peak is IR active but not Raman active in bulk $Bi_2Te_3$ [8, 19]. We attribute its appearance in FQL to breaking the crystal symmetry due to the presence of two interfaces. A single quintuple is inversely symmetric, which suggests that the identified crystal symmetry breaking is likely related to the loss of infinite crystal periodicity due to interfaces and corresponding relaxation of the phonon wave vector $q$=0 selection rule. Since $E_g^2$ (TO) is a regular BZ-center peak originating in the "bulk" of the film, the $I(A_{1u})/I(E_g^2)$ ratio increases with decreasing *H* because of the decreasing interaction volume *V*=*S*×*H* (*S* is the cross-sectional area of the laser spot), which defines $I(E_g^2)$ for *H* smaller than the light penetration depth in a given material. We estimated the light penetration depth in our samples to be 60 – 110 nm for 488-nm laser depending on the carrier concentration. This values correlates well with *H* when the 117 $cm^{-1}$ peak appears in FQL's spectrum.

Another intriguing observation from Fig. 3 (a) is an evolution of the $I(A^2_{1g})/I(E_g^2)$ ratio as one goes from bulk crystal to FQLs. In bulk, the measured intensity of in-plane vibrations $I(E_g^2)$ is higher than that of the out-of-plane vibrations $I(A^2_{1g})$, which is in agreement with literature for bulk $Bi_2Te_3$ [7-10]. But in the atomically thin films the intensity of the out-of-plane vibrations increases with decreasing thickness and at some *H* exceeds that of $I(E_g^2)$. In FQL with *H*=4 nm, the ratio $I(A^2_{1g})/I(E_g^2)>1$. At this point, we do not have quantitative explanation of the physics behind the observed trend. But it is reasonable to assume that the out-of-plane vibrations will be less restrained in a four-quintuple film than in bulk, which may lead to larger amplitudes of vibrations. This hypothesis is supported by comparison of the Raman spectra recorded from the suspended and supported portions of FQLs (inset to Fig. 1).

Table II provides frequencies of main peaks and the $I(A^2_{1g})/I(E_g^2)$ ratio for the suspended and supported FQLs. For the suspended FQL, the $I(A^2_{1g})/I(E_g^2)$ ratio is enhanced as compared to one for the supported FQL. The latter can be again attributed to the free-surface boundary conditions for the suspended FQL allowing for larger amplitude (intensity) of vibrations. Another possibility is an absence of the surface-charge effect, which is likely a factor influencing Raman peaks of FQL on $Si/SiO_2$ substrate. Another important observation from the data summarized in Table II is the fact that all peaks in the spectrum of suspended FQLs are red shifted by about ~2 $cm^{-1}$. We attribute it to bending of the suspended $Bi_2Te_3$ quintuples and resulting strain, which shifts the Raman peaks to smaller frequencies. Similar effects were observed in other strained materials [20]. Somewhat higher local heating effects in suspended FLQs may also lead to the red shift but very low excitation power bused in these experiments make this explanation less likely. The importance of the discussed modifications of Raman signatures of unstrained and strained $Bi_2Te_3$ films goes beyond TI research because bismuth telluride is also widely used as thermoelectric material [21].

In conclusion, we found that TO phonon mode $A_{1u}$, which is not-Raman active in bulk $Bi_2Te_3$ crystals, appears in the atomically-thin films due to crystal-symmetry breaking. The calibrated evolution of the $I(A_{1u})/I(E_g^2)$ and $I(A^2_{1g})/I(E_g^2)$ intensity ratios can be used for determining the number of quintuples in few-quintuple $Bi_2Te_3$ films produced for topological insulator and thermoelectric applications.

***Acknowledgements***

The authors acknowledge the support from DARPA – SRC through the FCRP Center on Functional Engineered Nano Architectonics (FENA) and Interconnect Focus Center (IFC) as well as from US AFOSR through contract A9550-08-1-0100.

**Table I**: Raman peaks in $Bi_2Te_3$ crystals and few-quintuple films

| | $E_g^1$ | $A^1_{1g}$ | $E_g^2$ | $A_{1u}$ | $A^2_{1g}$ | $I(A^2_{1g})/ I(E_g^2)$ | $I(A_{1u})/ I(E_g^2)$ | Comments |
|---|---|---|---|---|---|---|---|---|
| Bulk | 34.4 | 62.1 | 101.7 | - | 134.0 | 0.75 | - | This work |
| 82 nm | 35.8 | 61.5 | 101.9 | 116.9 | 132.7 | 0.83 | 0.62 | This work |
| 40 nm | 38.9 | 61.3 | 101.3 | 116.2 | 133.0 | 0.92 | 0.76 | This work |
| 4 nm | 38.9 | 60.9 | 101.4 | 116.7 | 132.9 | 1.30 | 0.86 | This work |
| Bulk | 36.5 | 62.0 | 102.3 | - | 134.0 | - | - | Ref. [7] |
| Bulk | - | 62.3 | 103.7 | - | 134.2 | - | - | Ref. [10] |

**Table II**: Raman spectra of few-quintuple $Bi_2Te_3$ films

| | $A^1_{1g}$ | $E_g^2$ | $A_{1u}$ | $A^2_{1g}$ | $I(A^2_{1g})/ I(E_g^2)$ |
|---|---|---|---|---|---|
| Supported | 62.9 | 103.0 | 119.5 | 134.2 | 0.93 |
| Suspended | 60.8 | 101.5 | 117.3 | 132.9 | 1.19 |

**FIGURE CAPTIONS**

Figure 1: SEM images of the mechanically exfoliated few-quintuple $Bi_2Te_3$ film with large lateral dimensions. The inset shows single-crystal bismuth telluride film suspended across a trench in $Si/SiO_2$ wafer.

Figure 2: (Color online) AFM measured profile of the exfoliated $Bi_2Te_3$ film showing a region with four quintuples. The inset is an AFM image of the scanned area with the red markers indicating the position of the tip, which correspond to the points in the height profile marked by the red arrows.

Figure 3: (Color online) (a) Raman spectra of reference bulk $Bi_2Te_3$ crystal and few-quintuple films. Note that $A_{1u}$ mode at ~117 $cm^{-1}$ becomes Raman active only in the thin films. (b) Schematic of the main lattice vibrations in {–$Te^{(1)}$–Bi–$Te^{(2)}$–Bi–$Te^{(1)}$–} quintuples. Large circles are Bi atoms. A small circle in the middle is $Te^{(2)}$ atom, a center for the inversion symmetry. The “+” and “-” signs in the schematics indicate atomic motions toward and from the observer. In all panels, the letters “E” and “A” indicate the in-plane and out-of-plane (*c* axis) lattice vibrations, respectively. The subscript “g” denotes Raman active while “u” stands for IR-active modes.

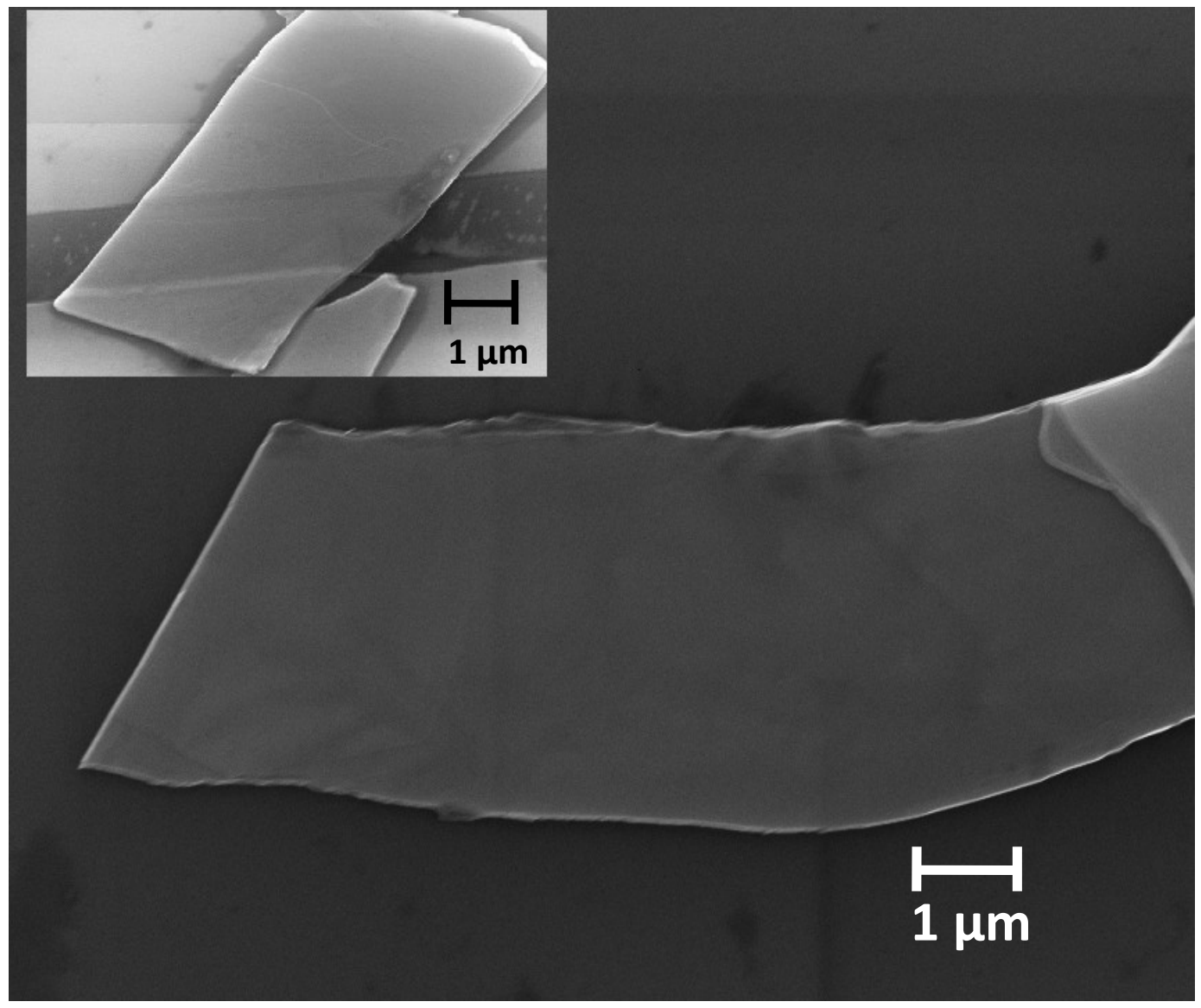

FIGURE 1

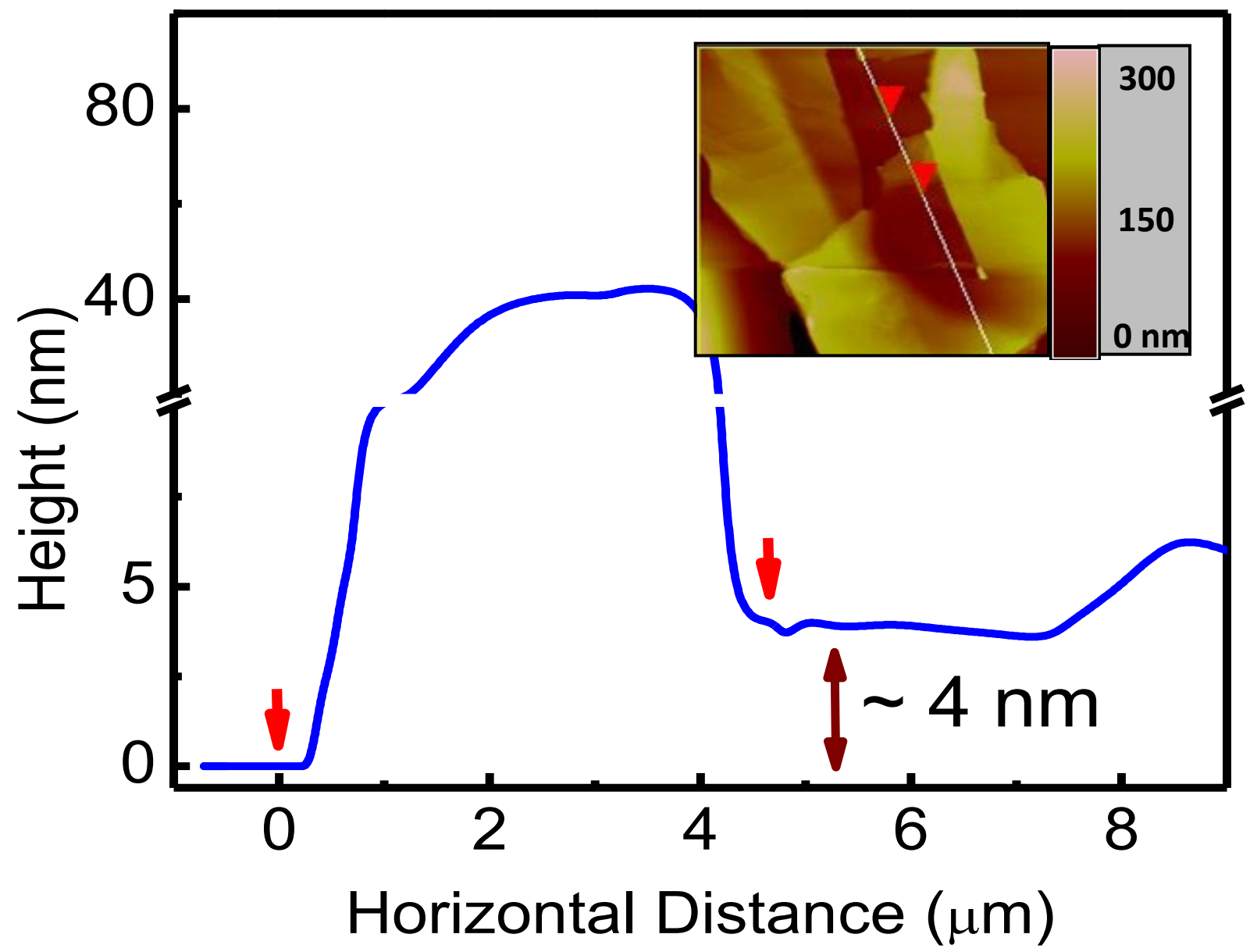
80
40
5
0
Height (nm)
0
2
4
6
8
Horizontal Distance (μm)
~ 4 nm
300
150
0 nm

FIGURE 2

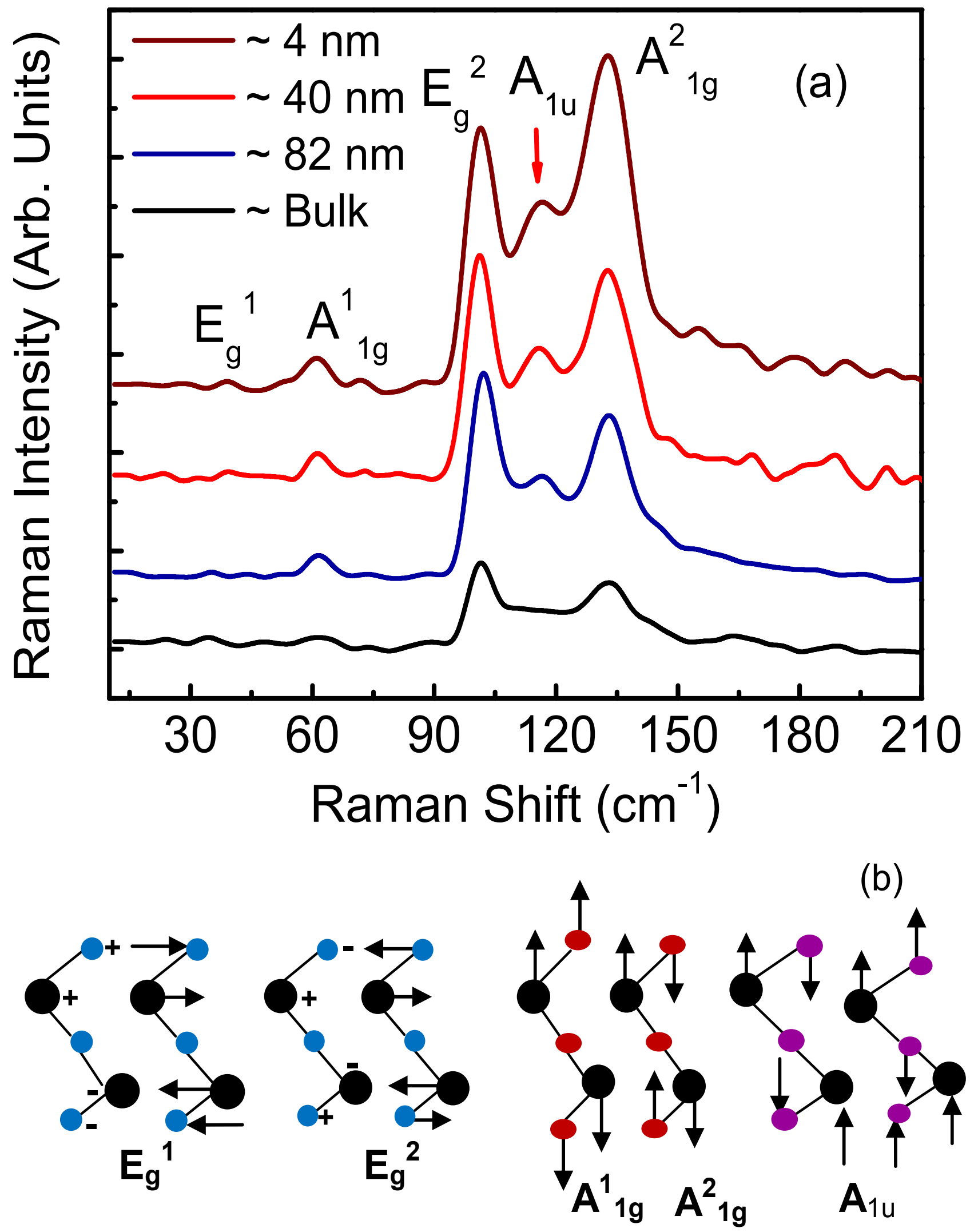
~ 4 nm
~ 40 nm
~ 82 nm
~ Bulk
E_g^2
A_1u
A^2_1g
(a)
E_g^1
A^1_1g
Raman Intensity (Arb. Units)
30
60
90
120
150
180
210
Raman Shift (cm^-1)
(b)
E_g^1
E_g^2
A^1_1g
A^2_1g
A_1u

FIGURE 3 (a, b)